\documentclass[%
 twocolumn,
 10pt,
superscriptaddress,
 amsmath,amssymb,
 aps,
 pra,
floatfix,
]{revtex4-2}

\usepackage{graphicx}
\usepackage{bm}
\usepackage{times}
\usepackage{amsmath,amssymb}
\usepackage{enumerate}
\usepackage{multirow}
\usepackage{qcircuit}

\usepackage[colorlinks=true,linkcolor=blue,citecolor=red, linktocpage=true,breaklinks=true]{hyperref}

\usepackage{epstopdf}

\newcommand{\eq}{\begin{equation}}
\newcommand{\en}{\end{equation}}
\newcommand{\eqa}{\begin{eqnarray}}
\newcommand{\ena}{\end{eqnarray}}
\newcommand{\tr}{\mathrm{Tr}}
\newtheorem{theorem}{Theorem}

\begin{document}

\title{Optimal realization of Yang-Baxter gate on quantum computers}

\author{Kun \surname{Zhang}}
\email{kunzhang@nwu.edu.cn}
\affiliation{School of Physics, Northwest University, Xi’an 710127, China}
\affiliation{Peng Huanwu Center for Fundamental Theory, Xi'an 710127, China}
\affiliation{Department of Chemistry, State University of New York at Stony Brook, Stony Brook, New York 11794, USA}
\author{Kwangmin \surname{Yu}}
\email{kyu@bnl.gov}
\affiliation{Computational Science Initiative, Brookhaven National Laboratory, Upton, New York 11973, USA}
\author{Kun \surname{Hao}}
\email{haoke72@163.com}
\affiliation{Peng Huanwu Center for Fundamental Theory, Xi'an 710127, China}
\affiliation{Institute of Modern Physics, Northwest University, Xi'an 710127, China}
\author{Vladimir \surname{Korepin}}
\email{vladimir.korepin@stonybrook.edu}
\affiliation{C.N. Yang Institute for Theoretical Physics, Stony Brook University, New York 11794, USA}

\date{\today}

\begin{abstract}

Quantum computers provide a promising method to study the dynamics of many-body systems beyond classical simulation. On the other hand, the analytical methods developed and results obtained from the integrable systems provide deep insights on the many-body system. Quantum simulation of the integrable system not only provides a valid benchmark for quantum computers but is also the first step in studying integrable-breaking systems. The building block for the simulation of an integrable system is the Yang-Baxter gate. It is vital to know how to optimally realize the Yang-Baxter gates on quantum computers. Based on the geometric picture of the Yang-Baxter gates, we present the optimal realizations of two types of Yang-Baxter gates with a minimal number of CNOT or $R_{zz}$ gates. We also show how to systematically realize the Yang-Baxter gates via the pulse control. We test and compare the different realizations on IBM quantum computers. We find that the pulse realizations of the Yang-Baxter gates always have a higher gate fidelity compared to the optimal CNOT or $R_{zz}$ realizations. On the basis of the above optimal realizations, we demonstrate the simulation of the Yang-Baxter equation on quantum computers. Our results provide a guideline and standard for further experimental studies based on the Yang-Baxter gate.

	
\end{abstract}

\maketitle

\section{\label{sec:intro} Introduction}

The Yang-Baxter equation plays the central role in the study of exactly solvable or integrable models \cite{Perk2006,jimbo1990yang}. It was independently introduced by C. N. Yang and R. J. Baxter in solving quantum field models and statistical models, respectively \cite{yang1967some,yang1968s,baxter1972partition,baxter2016exactly}. The Leningrad school of Faddeev constructed a systematic method to solve integrable quantum systems, called the quantum inverse scattering method, where the Yang-Baxter equation is the starting point in this method \cite{takhtadzhan1979quantum,korepin1997quantum}.

In the last twenty years, the Yang-Baxter equation has been introduced into the study of quantum information and computation. Unitary representations of the braid group relation, which gives the asymptotic form of the Yang-Baxter equation, reveal the relationship between the topological entanglement and the quantum entanglement \cite{kauffman2004braiding,padmanabhan2020quantum,padmanabhan2020generating}. Then through a mathematical procedure, called Yang-Baxterization, the unitary braid group representation can generate the unitary solutions of the Yang-Baxter equation, called the Yang-Baxter gate \cite{zhang2005universal,zhang2005yang}. The two-qubit Yang-Baxter gate can generate arbitrary degrees of entanglement. Therefore, the Yang-Baxter gate with single-qubit gates can form the universal quantum gate set and realize various quantum information processing protocols \cite{chen2007braiding,wang2009entanglement,yu2014factorized,ge2016yang}.

In another aspect, the quantum simulation circuit of the integrable system based on Trotter decomposition preserves the integrability, called the integrable circuit \cite{vanicat2018integrable}. The building block of the integrable circuit is the Yang-Baxter gate, and the transfer matrix can be obtained based on the Yang-Baxter equation \cite{miao2022floquet}. Similarly to the integrable system, the correlation function of the integrable circuit can be efficiently computed \cite{claeys2022correlations}. The integrable circuit not only represents the simulation of the integrable system, but also serves as a perfect tool to benchmark large-scale quantum computers because of its exactly solvable nature. Recently, the integrable circuit, composed of the Yang-Baxter gate, was implemented on IBM quantum computers, and the conserved charges of the circuit are measured \cite{maruyoshi2023conserved}. In addition, the propagation of bound states has recently been observed in the integrable circuit \cite{morvan2022formation}. 

Because of the importance of Yang-Baxter gate and Yang-Baxter equation, many researches have studied how to realize and implement the Yang-Baxter gate and Yang-Baxter equation on quantum computers.  For example, the Yang-Baxter gates from the integrable circuit have been implemented on IBM superconducting-based quantum computers \cite{maruyoshi2023conserved,peng2022quantum}; the Yang-Baxter equation has been tested on the optical and NMR systems \cite{hu2008optical,zheng2013direct,vind2016experimental,zheng2018duality,wang2020experimental}. 

However, the problem of how to optimally realize the Yang-Baxter gates is never addressed. Here, the optimal realization means the most efficient decompositions of Yang-Baxter gates into the fundamental operations of quantum computers. In general cases, the less gate implementation time gives the higher gate fidelity. The realizations can be categorized into gate-based or pulse-based realizations. The optimal gate-based realization is to decompose arbitrary gates into as few as possible native or pre-defined gates of quantum computers \cite{zhang2004optimal,vidal2004universal,vatan2004optimal,shende2004minimal}. The pulse-based realization is to construct arbitrary gates with pulse operations applied on the qubits \cite{alexander2020qiskit}. Here the pulse-based realization represents a lower level control of quantum computers compared to the gate-based realization. The flexibility of pulse control provides many advantages, as demonstrated in recent studies \cite{gokhale2020optimized,stenger2021simulating,earnest2021pulse,niu2022effects,satoh2022pulse,ibrahim2022pulse,chen2022error}. 

In this work, we design and compare different realizations of two types of Yang-Baxter gates on quantum computers. One Yang-Baxter gate is from the construction of unitary solutions of the Yang-Baxter equation \cite{zhang2005universal,zhang2005yang}; the other is from the integrable circuit \cite{vanicat2018integrable,maruyoshi2023conserved}. We find that pulse-based realizations of Yang-Baxter gates with different spectral parameters always give the shortest implementation time and, therefore, the highest gate fidelity. The Yang-Baxter equation from different realizations of the Yang-Baxter gates is also demonstrated in our study.  

Our paper is organized as follows. In Sec. \ref{sec:YBG}, we review the Yang-Baxter equation and the Yang-Baxter gates. In Sec. \ref{sec:Realization_YBG}, we show how to design different realizations of Yang-Baxter gates. We test these realizations on quantum computers. In Sec. \ref{sec:sim_YBE}, we study the simulation of the Yang-Baxter equation based on different realizations of the Yang-Baxter gates. Sec. \ref{sec:conclusion} is the conclusion. Appendix \ref{App:A} includes some details on the construction of the Yang-Baxter gate. Appendix \ref{App:B} presents details on the IBM quantum processors used in this study.

\section{\label{sec:YBG} Yang-Baxter equation and Yang-Baxter gate}

In this section, we first review the Yang-Baxter equation and the braid group relation in Sec \ref{subsec:YBE}. Then we introduce two types of Yang-Baxter gates in Secs. \ref{subsec:YBG_Baxterization} and \ref{subsec:YBG_integrable_circuit}, which originate from different studies. 

\subsection{\label{subsec:YBE}Yang-Baxter equation}

\begin{figure}
    \begin{center}
    	\includegraphics[width=\columnwidth]{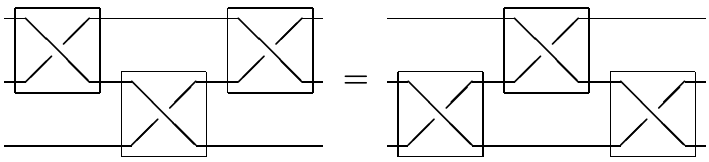}
    	\caption{Yang-Baxter equation in the quantum circuit model. The crossing represents the solution of the Yang-Baxter equation. The box represents the two-qubit gate. }\label{fig_YBE}
    \end{center}
\end{figure}

The Yang-Baxter equation is a consistency relation with respect to the two-body scattering matrix \cite{Perk2006,jimbo1990yang}. It says that the two-body scattering matrix in the three-body interaction can be interchangeable. Consider the qubit or spin-1/2 system. Then the
$4\times 4$ two-body scattering matrix $\check R$ acts on the tensor product space $\mathcal H_2\otimes \mathcal H_2$. The Yang-Baxter equation with an additive parameter reads as
\begin{multline}
\label{eq:YBE_add}
    \check R_{12}(\mu_1)\check R_{23}(\mu_1+\mu_2)\check R_{12}(\mu_2) \\
    = \check R_{23}(\mu_2)\check R_{12}(\mu_1+\mu_2)\check R_{23}(\mu_1).
\end{multline}
Here, $\mu$ is also called the spectral parameter, which is related to the scattering momentum. Taking the logarithmic parameter $u = \log \nu$, we can have the Yang-Baxter equation with the multiplicative parameter, given by 
\begin{equation}
\label{eq:YBE_multi}
    \check R_{12}(\nu_1)\check R_{23}(\nu_1\nu_2)\check R_{12}(\nu_2) = \check R_{23}(\nu_2)\check R_{12}(\nu_1\nu_2)\check R_{23}(\nu_1).
\end{equation}
See FIG. \ref{fig_YBE} for the graphic representation of the Yang-Baxter equation. Note that the $\check R$ matrix is related to the Hamiltonian density of the system.

It is well-known that the Yang-Baxter equation asymptotically reduces to the braid group representation \cite{kauffman2001knots}. The braid group has the generator $b_j$ with $j=1,2,\ldots,n-1$ satisfying 
\begin{subequations}
\begin{align}
    & b_jb_{j\pm 1}b_j =  b_{j\pm 1}b_j  b_{j\pm 1}, \\
    & b_jb_k = b_kb_j,\qquad |j-k|\geq 2.
\end{align}
\end{subequations}
The braid group relation is also called the parameter-independent Yang-Baxter equation or the constant Yang-Baxter equation.

\subsection{\label{subsec:YBG_Baxterization}Yang-Baxter gate from Yang-Baxterization}

The braid group relation or the constant Yang-Baxter equation characterizes the low-dimensional topology \cite{kauffman2001knots}. The topological entanglement can be described by the corresponding knot invariants. Then one can consider the unitary representation of the braid group. It works as a two-qubit gate, which can generate quantum entanglement. See FIG. \ref{fig_YBE}. We call it as the parameter-independent Yang-Baxter gate or the braid gate. How to construct the unitary braid gate is well studied \cite{dye2003unitary,padmanabhan2020braiding}.

Furthermore, the braid group representation can generate the solution of Yang-Baxter equation through the procedure called the Yang-Baxterization \cite{jimbo1985q,jimbo1986q}. The Yang-Baxter gate generated from the braid gate has been extensively studied \cite{zhang2005universal,zhang2005yang}. At the two-qubit basis $\{|00\rangle,|01\rangle,|10\rangle,|11\rangle\}$, consider a typical braid gate 
\begin{equation}
\label{eq:braid_gate}
    \check B = \frac 1 {\sqrt 2}\begin{pmatrix}
    1 & 0 & 0 & 1 \\
    0 & 1 & -1 & 0 \\
    0 & 1 & 1 & 0 \\
    -1 & 0 & 0 & 1  \\
    \end{pmatrix},
\end{equation}
which gives the braid group representation $b_j = 1\!\!1^{\otimes j-1}\otimes \check B\otimes 1\!\!1^{\otimes n-j-1}$. The braid gate $\check B$ can generate the Bell state from the product state. For example $\check B|00\rangle = \left(|00\rangle-|11\rangle\right)/\sqrt 2$. The above two-qubit braid gate can be generalized into multiqubit cases via the extraspecial 2-groups \cite{Rowell2010extraspecial}. 

The above braid gate $\check B$ has two distinct eigenvalues $\lambda_{\pm} = (1\pm i)/\sqrt 2$. The Yang-Baxterization gives
\begin{equation}
    \check R_I(\nu) = \frac{1}{\sqrt{\mathcal N}}\left(\check B + \nu \lambda_+\lambda_- \check B^\dag\right).
\end{equation}
Here $\mathcal N = (1+\nu^2)$ is the normalization required by $\check R_I(\nu)\check R^\dag_I(\nu) = 1\!\!1_4$. The unitary condition also requires that the spectral parameter $\nu$ is real. When $\nu=0$, it automatically returns to the braid gate. In our study, we call it the type-I or first type Yang-Baxter gate. One can verify that $\check R_I(\nu)$ satisfies the Yang-Baxter equation with the multiplicative parameter, given in Eq. (\ref{eq:YBE_multi}). The Yang-Baxterization procedure may not give the matrix satisfying the Yang-Baxter equation if the braid gate has more than two distinct eigenvalues \cite{zhang2005yang}.

If we redefine the parameter as $\tan(\theta - \pi/4) = \nu$, then we have 
\begin{equation}
\label{eq:YBG_first}
    \check R_I(\theta) = \begin{pmatrix}
    \cos\theta & 0 & 0 & \sin\theta \\
    0 & \cos\theta & -\sin\theta & 0 \\
    0 & \sin\theta & \cos\theta & 0 \\
    -\sin\theta & 0 & 0 & \cos\theta  \\
    \end{pmatrix}.
\end{equation}
Clearly, when $\theta = \pi/4$, we have $\check R_I(\theta = \pi/4) = \check B$; when $\theta = 0$, we have $\check R_I(\theta = 0) = 1\!\!1_4$.

\subsection{\label{subsec:YBG_integrable_circuit}Yang-Baxter gate from integrable circuit}

In another research direction, the Yang-Baxter gate appears in the quantum simulation of the integrable spin chain \cite{vanicat2018integrable}. Consider the Heisenberg XXX model, with the Hamiltonian \cite{korepin1997quantum}
\begin{equation}
    H_{XXX} = \frac J 2 \sum_{j=1}^{2N} \left(1+\vec\sigma_j\cdot \vec\sigma_{j+1}\right).
\end{equation}
Here $\vec\sigma_j = \left(\sigma_x,\sigma_y,\sigma_z\right)$ is the Pauli vector with the Pauli matrices $\sigma_{x,y,z}$. The interaction can be separated into terms acting on the even or odd site. Then apply the Trotter decomposition to the evolution operator $e^{-iH_{XXX}t}$. The building block to simulate the dynamics of the Heisenberg XXX model is the following two-qubit gate
\begin{equation}
\label{eq:R_II_P}
    \check R_{II}(\mu) = \frac{1+i\mu P}{1+i\mu},
\end{equation}
with the two-qubit permutation operator $P = \left(1+\vec\sigma_j\cdot\vec\sigma_{j+1}\right)/2$. It is also called the SWAP gate in quantum computation \cite{nielsenQuantumComputationQuantum2010}. When $\mu\rightarrow\infty$, we get the permutation operator, which also gives the representation of the braid group. Note that the Hamiltonian density of the Heisenberg XXX model is also described by the permutation operator $P$. It also gives the representation of the braid group. One can verify that $\check R_{II}(\mu)$ satisfies the Yang-Baxter equation with the additive parameter, given in Eq. (\ref{eq:YBE_add}). We call it the type-II or second type Yang-Baxter gate. 

If we parameterize the spectral parameter as $\tan\phi = \mu$, up to a overall factor, we have 
\begin{equation}
\label{eq:YBG_second}
    \check R_{II}(\phi) \cong \begin{pmatrix}
    e^{i\phi} & 0 & 0 & 0 \\
    0 & \cos\phi & i\sin\phi & 0 \\
    0 & i\sin\phi & \cos\phi & 0 \\
    0 & 0 & 0 & e^{i\phi}  \\
    \end{pmatrix}.
\end{equation}
Clearly, when $\phi = \pi/2$, we have the SWAP gate $\check R_{II}(\phi = \pi/2) = iP$; when $\phi = 0$, we have $\check R_{II}(\phi = 0) = 1\!\!1_4$. The parameter $\phi$ is related to the Trotterized time step $n$ and the evolution time, namely $\phi = -tJ/n$. The Trotter evolution circuit given by $\check R_{II}(\phi)$ is also integrable, therefore called the integrable circuit \cite{vanicat2018integrable}. Because of the staggered structure of the circuit, the conserved charge of the circuit is different with the Heisenberg XXX model \cite{maruyoshi2023conserved}. Only in the Trotter time limit $n\rightarrow\infty$, it is identical to the evolution of the XXX model.

\section{\label{sec:Realization_YBG} Realizations of the Yang-Baxter gate}

Firstly, we review the general theories about the two-qubit gate realization in Sec. \ref{subsec:two_qubit_decompose}. Then we introduce the pulse realization of superconducting quantum computers in Sec. \ref{subsec:pulse_realization}. We present the realizations of two types of Yang-Baxter gates in Secs. \ref{subsec:YBG_first} and \ref{subsec:YBG_second}. The Yang-Baxter gate properties are also studied in details. 

\subsection{\label{subsec:two_qubit_decompose}Two-qubit gate properties and realizations}

In classical reversible computation, one- and two-bit gates are not universal. It means that the three-bit gate, such as the Toffoli or the Fredkin gate, is needed to do arbitrary computation \cite{toffoli1980reversible}. Remarkably, quantum computation only needs one- and two-qubit gates \cite{divincenzo1995two}. The most common two-qubit gate in quantum computation is the controlled-NOT (CNOT) gate, defined as \cite{nielsenQuantumComputationQuantum2010}
\begin{equation}
    \text{CNOT} = |0\rangle\langle 0|\otimes 1\!\!1_2 + |1\rangle\langle 1|\otimes\sigma_x,
\end{equation}
where the first qubit is the controlled qubit and the second qubit is the target qubit. It is well-known that CNOT gate with the single-qubit gates are universal \cite{barenco1995elementary}. A two-qubit gate is called universal if it can do the universal computation supplemented with the single-qubit gates. It turns out that any two-qubit gate that can generate entanglement from the product state is universal \cite{brylinski2002universal,bremner2002practical}. 

Another commonly used two-qubit gate is the Ising coupling gate given by
\begin{equation}
    R_{zz}(\theta) = e^{-\frac{i}{2}\theta\sigma_z\otimes\sigma_z}.
\end{equation}
It is a natural native two-qubit gate for trapped-ion quantum computers \cite{bruzewicz2019trapped}. One can also define the gate $R_{xx}$ or $R_{zx}$ correspondingly. CNOT gate and the $R_{zz}(\theta)$ gate with $\theta\neq k\pi$ and the integer $k$ can be converted to each other. A very useful gate identity is
\begin{equation}
\label{eq:Rz_two_cnot}
    \Qcircuit @C=0.7em @R=1.5em {
    &&&&& \ctrl{1} & \qw & \ctrl{1} & \qw \\
    \raisebox{0.8cm}{$R_{zz}(\theta)=$} &&&&& \targ & \gate{R_z(\theta)} & \targ & \qw 
    }
\end{equation}
with the single-qubit rotation gate $R_z(\theta)=e^{-\frac{i}{2}\theta\sigma_z}$. 

\begin{figure}
    \begin{center}
    	\includegraphics[width=\columnwidth]{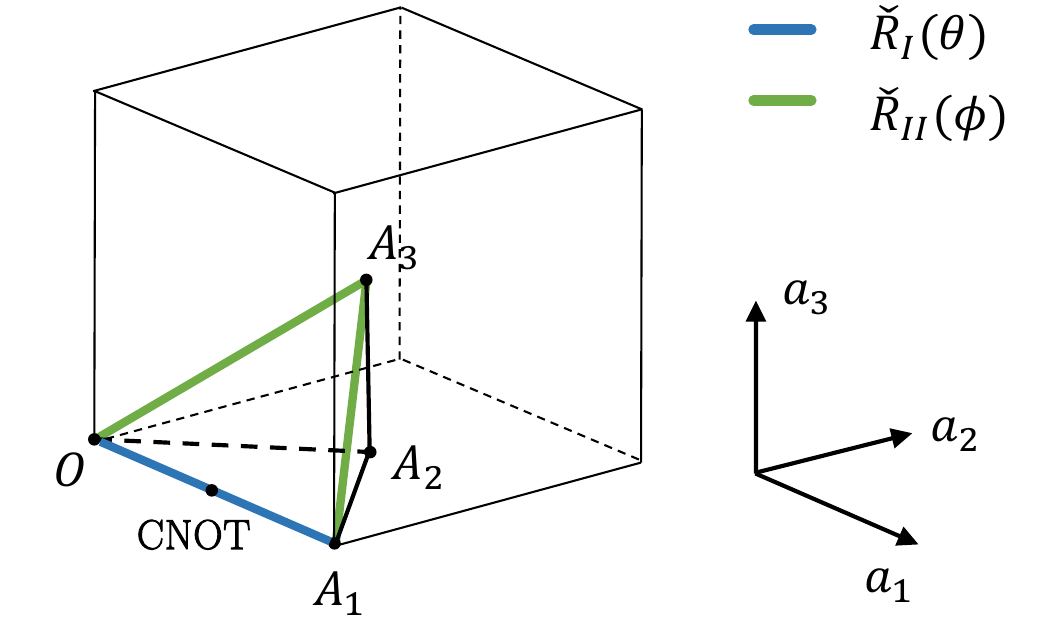}
    	\caption{Geometric representation of two-qubit gate $U=[a_1,a_2,a_3]$. The tetrahedron $OA_1A_2A_3$ has the vertexes $O=[0,0,0]$, $A_1=[\pi,0,0]$, $A_2=[\pi/2,\pi/2,0]$, and $A_3=[\pi/2,\pi/2,\pi/2]$. The Yang-Baxter gates $R_I(\theta)$ and $R_{II}(\phi)$ correspond to the blue and green edges respectively.} \label{fig_two-qubit}
    \end{center}
\end{figure}

Any two-qubit gate $U\in SU(4)$ has the decomposition \cite{kraus2001optimal}
\begin{equation}
\label{eq:U_decomposition}
    U = (V_1\otimes V_2)e^{\frac i 2\left(a_1\sigma_x\otimes \sigma_x + a_2\sigma_y\otimes \sigma_y + a_3\sigma_z\otimes \sigma_z\right)}(V_3\otimes V_4),
\end{equation}
with the single-qubit gate $V_k\in SU(2)$ ($k=1,2,3,4$). Therefore, only $[a_1,a_2,a_3]$, called the nonlocal parameters, characterizes the entanglement properties of $U$. Two-qubit gates with the same nonlocal parameters are equivalent in terms of the single-qubit gates. Limit to the parameter region $\pi-a_2\geq a_1\geq a_2\geq a_3\geq 0$, there is nice geometric representation of two-qubit gate. Specifically, every two-qubit gate is one-to-one correspondence with the point contained in a tetrahedron except points on the $OA_1A_2$ base \cite{zhang2003geometric}. See FIG. \ref{fig_two-qubit}. The middle of $OA_1$ represents the CNOT gate with $[\text{CNOT}] = [\pi/2,0,0]$. The $R_{zz}$ gate is at the line $OA_1$ with $[R_{zz}(\theta)] = [\theta,0,0]$. The SWAP gate defined in Eq. (\ref{eq:R_II_P}) is at the point $A_3$ with $[P] = [\pi/2,\pi/2,\pi/2]$. 

The entangling power of quantum gate can quantify its ability of generating entanglement from the product state. The entangling power of two-qubit gate is defined as \cite{zanardi2000entangling}
\begin{equation}
    e_p(U) = \overline{E(U|\psi_1\rangle\otimes|\psi_2\rangle)}_{|\psi_1\rangle\otimes|\psi_2\rangle},
\end{equation}
with the linear entropy of two-qubit state $E(|\Psi\rangle) = 1 - \tr_1\rho^2$ and $\rho = \tr_2|\Psi\rangle\langle\Psi|$. The over line means taking the average over all product states $|\psi_1\rangle\otimes|\psi_2\rangle$ with the uniform distribution. Because of the simple decomposition given in Eq. (\ref{eq:U_decomposition}), one can find that the entangling power of the two-qubit gate is \cite{balakrishnan2010entangling}
\begin{align}
\label{eq:e_p}
    e_p(U) = \frac 2 9 (1 - G(U)),
\end{align}
with the local invariants of the two-qubit gate
\begin{equation}
    G(U) = \cos^2a_1\cos^2a_2\cos^2a_3 + \sin^2a_1\sin^2a_2\sin^2a_3.
\end{equation}
The CNOT gate with the nonlocal parameter $[\text{CNOT}] = [\pi/2,0,0]$ has the maximal entangling power $e_p(\text{CNOT}) = 2/9$. The SWAP gate with $[P] = [\pi/2,\pi/2,\pi/2]$ has zero entangling power. The gate $[R_{zz}(\theta)] = [\theta,0,0]$ has the entangling power $2\sin^2\theta/9$. Moreover, we can see that any two-qubit gate except the SWAP gate has non-zero entangling power. Therefore, they are all universal. 

Entangling power only characterizes the average entanglement generated from that gate. Although few two-qubit gates have the maximal entangling power, this does not mean that other two-qubit gates can not generate the maximal entangled state from the product state. Based on the geometric picture of the two-qubit gate, it has been proven \cite{zhang2003geometric}
\begin{theorem}
\label{theorem_generate_maximal}
    A two-qubit gate with the nonlocal parameters $[U] = [a_1,a_2,a_3]$ can generate the maximal entangled state from the product state iff
    \begin{equation}
    \label{eq:perfect_entangler_1}
        \frac \pi 2 \leq a_j+a_k \leq a_j+a_l+\frac \pi 2 \leq \pi,
    \end{equation}
    or 
    \begin{equation}
    \label{eq:perfect_entangler_2}
        \frac {3\pi} 2 \leq a_j+a_k \leq a_j+a_l+\frac \pi 2 \leq 2\pi.
    \end{equation}
    Here $(j,k,l)$ is the permutation of $(1,2,3)$.
\end{theorem}
It is well known that the Bell state can be generated from CNOT gate, which satisfies condition (\ref{eq:perfect_entangler_1}). Only when the $\theta=\pi/2$ or $\theta = 3\pi/2$, $R_{zz}(\theta)$ gate can generate the maximal entangled state, which is also local equivalent to the CNOT gate. 

How to construct arbitrary two-qubit gate from minimal numbers of CNOT gates is well studied \cite{zhang2004optimal,vidal2004universal,vatan2004optimal,shende2004minimal}. At most three CNOTs are required. When $U\in SO(4)$ or equivalently with vanishing nonlocal parameters $[U] = [a_1,a_2,0]$, only two CNOTs are needed. If we want to construct two-qubit gates from other two-qubit gate, such as $R_{zz}$ gate, a very useful criterion is \cite{zhang2004optimal}
\begin{theorem}
\label{theorem_minimal_rzz}
    Any two-qubit gate $[U] = [a_1,a_2,a_3]$ can be constructed from minimal $n$ ($n\geq 3$) applications of $R_{zz}(\theta)$ gate, if the nonlocal parameters satisfies $0\leq a_1+a_2+a_3\leq n\theta$ or $a_1-a_2-a_3\geq \pi-n\theta$.
\end{theorem}
The above theorem also implies that at most three CNOT gates can realize any two-qubit gate.

\subsection{\label{subsec:pulse_realization}Pulse realization of the two-qubit gate}

Quantum gates are the building blocks for quantum computation. However, they are not the lowest-level control from the view point of physics. The dynamics of qubits is controlled by pulses. For example, IBM quantum computers with superconducting transmon qubits realize the CNOT gate via the so-called cross-resonance operation \cite{chow2011simple,alexander2020qiskit}. The cross-resonance operation has the two-qubit effective Hamiltonian \cite{alexander2020qiskit}
\begin{equation}
H_{CR} = \frac{\sigma_z\otimes C}{2} + \frac{1\!\!1_2\otimes D}{2},
\end{equation}
with $C = \omega_{z1\!\!1}1\!\!1_2 + \omega_{zx}\sigma_x + \omega_{zy}\sigma_y + \omega_{zz}\sigma_z$ and $D = \omega_{1\!\!1x}\sigma_x + \omega_{1\!\!1y}\sigma_y + \omega_{1\!\!1z}\sigma_z$. To realize the CNOT gate from the cross-resonance operation $U_{CR} = e^{-it_{CR}H_{CR}}$, unwanted terms in $H_{CR}$ must be canceled. The goal is to isolate the $\sigma_z\otimes\sigma_x$ term and then to realize the $R_{zx}(\pi/2)$ gate, which is local equivalent to the CNOT gate. 

\begin{figure}[t!]
    \begin{center}
    	\includegraphics[width=\columnwidth]{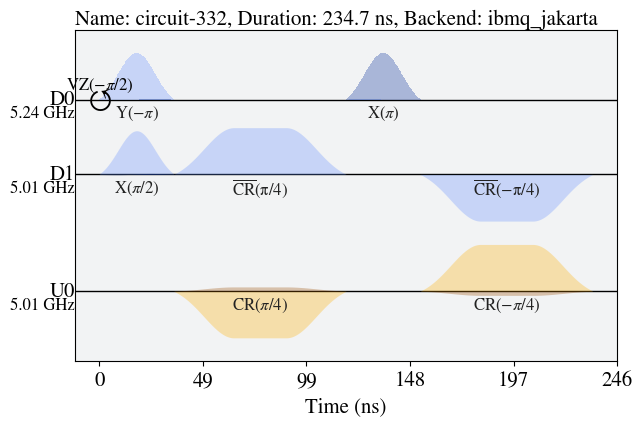}
    	\caption{Pulse realization of CNOT gate on qubit 0 and 1 of IBM {\fontfamily{qcr}\selectfont ibmq\_jakarta} processor. The D0 and D1 are the {\fontfamily{qcr}\selectfont DriveChannel} for the single-qubit gate on qubit 0 and 1. The U0 is the {\fontfamily{qcr}\selectfont ControlChannel} for the two-qubit interaction. The circular arrow represents the virtual single-qubit $z$-axis rotation gate. Pulses with the bright and dark colors represent the in-phase and quadrature-phase components. The negative amplitude means that the pulse has a $\pi$-phase.} \label{fig_cnot}
    \end{center}
\end{figure}

Echoed cross-resonance pulses with opposite phases, combined with echoed $\pi$ pulses, are applied to suppress the $1\!\!1_2\otimes\sigma_x$ and $\sigma_z\otimes 1\!\!1_2$ terms \cite{corcoles2013process}. In addition, compensation pulses on the target qubit are used to reduce the errors of terms $1\!\!1_2\otimes\sigma_y$ and $\sigma_z\otimes \sigma_z$ \cite{sundaresan2020reducing}. Moreover, additional pulses of single-qubit gates are required to convert $R_{zx}(\pi/2)$ to the CNOT gate. Note that the single-qubit rotation gate around the $z$-axis is realized virtually by tracking the pulse phase \cite{mckay2017efficient}. See FIG. \ref{fig_cnot} for the demonstration of the pulse realization of the CNOT gate on qubits 0 and 1 of IBM {\fontfamily{qcr}\selectfont ibmq\_jakarta} processor. Note that the cross-resonance operation is realized by a Gaussian-square shape pulse.

\begin{figure}[t!]
    \begin{center}
    	\includegraphics[width=\columnwidth]{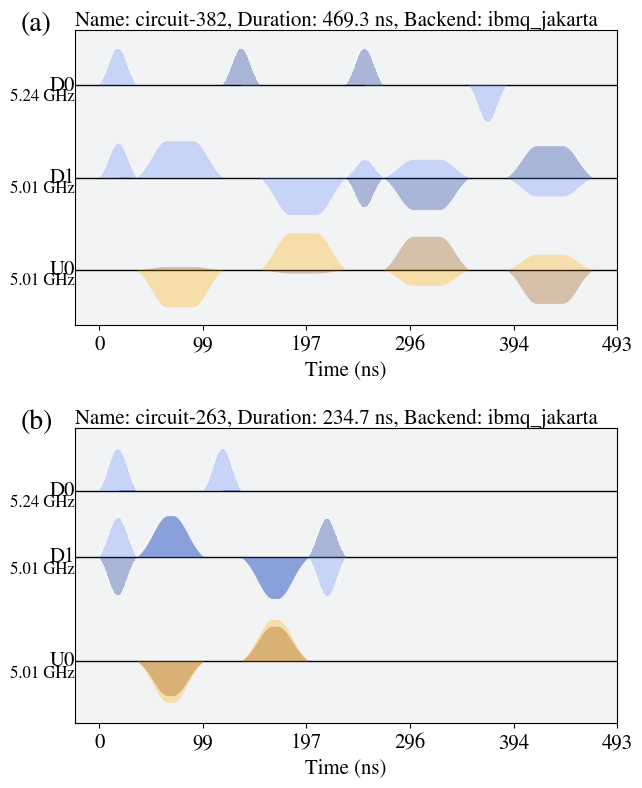}
    	\caption{The pulses of $R_{zz}(\pi/3)$ gate realized from (a) two CNOTs conjugated on the $R_z(\pi/3)$ gate as shown in Eq. (\ref{eq:Rz_two_cnot}); one echo cross-resonance operations with rescaled parameters of CNOT gate. The virtual $z$-axis rotation is omitted here for simplicity. } \label{fig_rzz}
    \end{center}
\end{figure}

Through the cross-resonance operation, the pulse control has the ability to realize any two-qubit gates. However, finding the optimal pulse for specific two-qubit gates requires heavy calibrations, which is time-consuming. To bypass the time-consuming calibrations, one can use the calibrated parameters of the CNOT gate to infer and construct the other two-qubit gate. Specifically, if the area under the cross-resonance pulse changes, it becomes $R_{zx}(\theta)$ instead of the fixed $\theta = \pi/2$ for the CNOT gate \cite{gokhale2020optimized,stenger2021simulating,earnest2021pulse}. The Gaussian square pulse has the area
$$
\alpha = |A|\left(\omega + \sigma\sqrt{2\pi}\text{erf}(n_\sigma)\right),
$$
with the amplitude $A$, flat-top width $\omega$, standard deviation $\sigma$ of the Gaussian tails, and the truncated number $n_\sigma$. Then the gate $R_{zx}(\theta)$ with $\theta\leq\pi/2$ has the target area $\alpha(\theta)$ given by $\alpha(\theta)/\alpha(\pi/2) = 2\theta/\pi$. If the Gaussian part of the $R_{zx}(\pi/2)$ pulse is smaller than $\alpha(\theta)$, then we cut part of the flat-top pulse (decreasing $\omega$). If the Gaussian part does not have enough area, then the amplitude is rescaled accordingly (decreasing $A$). As an example, the $R_{zz}(\theta)$ gate can be realized from either two CNOTs given by Eq. (\ref{eq:Rz_two_cnot}), or directly through the cross-resonance operation. See FIG. \ref{fig_rzz} for comparison of pulses. The pulse duration of the rescaled cross-resonance operation is only half of the two CNOTs pulses. Consequently, gate fidelity improves, as demonstrated in \cite{stenger2021simulating,earnest2021pulse,chen2022error}.

\subsection{\label{subsec:YBG_first}Realizations of first type of Yang-Baxter gate}

\begin{figure}
    \begin{center}
    	\includegraphics[width=\columnwidth]{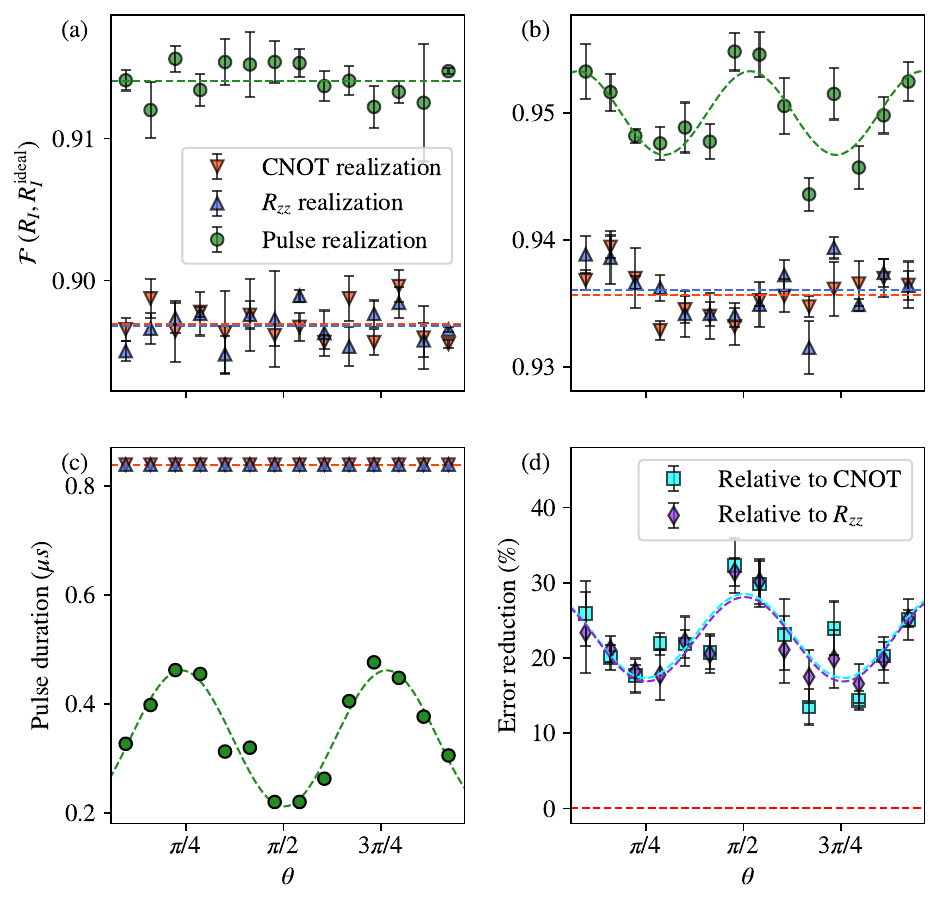}
    	\caption{Realizations of Yang-Baxter gate $\check R_I(\theta)$ on the quantum computer {\fontfamily{qcr}\selectfont ibmq\_montreal}. Measured gate fidelity on (a) FakeMontreal ({\fontfamily{qcr}\selectfont ibmq\_montreal} simultor) or (b) {\fontfamily{qcr}\selectfont ibmq\_montreal}. (c) The pulse durations of three different realizations. (d) The error reduction of pulse realization compared to the CNOT or $R_{zz}$ realization. The dashed line is the fitting. Each tomography circuit repeats four times and each time runs 4096 shots. }\label{fig_YBG_fidelity_1}
    \end{center}
\end{figure}

The first type Yang-Baxter gate $\check R_I(\theta)$ defined in Eq. (\ref{eq:YBG_second}) is an element of the $SO(4)$ group. According to \cite{vidal2004universal,vatan2004optimal}, at most two CNOTs are needed to realize it. Specifically, we have
\begin{equation}
\label{eq:R_I_two_CNOTs}
    \Qcircuit @C=0.6em @R=1.5em {
    &&&&& \gate{S} & \gate{H} & \ctrl{1} & \qw & \ctrl{1} & \gate{H} & \gate{S^\dag} & \qw \\
    \raisebox{1.2cm}{$\check R_{I}(\theta)=$} &&&&& \gate{H} & \qw & \targ & \gate{R_z(2\theta)} & \targ & \gate{H} & \qw & \qw
    }
\end{equation}
with the Hadamard gate $H = (\sigma_x+\sigma_z)/\sqrt{2}$ and the phase gate $S=\text{diag}(1,i)$. Compare the above decomposition with the realization of $R_{zz}(\theta)$ gate given by Eq. (\ref{eq:Rz_two_cnot}), then we know 
\begin{equation}
\label{eq:R_I_one_Rzz}
    \Qcircuit @C=0.6em @R=1.5em {
    &&&&& \gate{S} & \gate{H} & \multigate{1}{R_{zz}(2\theta)} & \gate{H} & \gate{S^\dag} & \qw \\
    \raisebox{1.2cm}{$\check R_{I}(\theta)=$} &&&&& \gate{H} & \qw & \ghost{R_{zz}(2\theta)} & \gate{H} & \qw & \qw
    }
\end{equation}
The $\check R_I(\theta)$ gate is local equivalent to the $R_{zz}(2\theta)$ gate. Then we know that $\check R_I(\theta)$ and $R_{zz}(2\theta)$ have the same nonlocal parameters $[\check R_I(\theta)] = [R_{zz}(2\theta)] = [2\theta,0,0]$. In the geometric representation of the two-qubit gate \cite{zhang2003geometric}, the $\check R_I(\theta)$ gate corresponds to the line $OA_1$. See FIG. \ref{fig_two-qubit}. When $\theta = \pi/4$, the $\check R_I(\pi/4)$ gate is local equivalent to the CNOT gate. Specifically, we have
\begin{equation}
    \Qcircuit @C=0.7em @R=1.5em {
    &&&&& \gate{H} & \gate{S^\dag} & \ctrl{1} & \gate{H} & \qw & \qw & \qw \\
    \raisebox{1.2cm}{$\check R_{I}(\pi/4)=$} &&&&& \gate{S^\dag} & \qw & \targ & \gate{S} & \gate{H} & \gate{\sigma_z} & \qw 
    }
\end{equation}
which is equivalent to the gate realization reported in \cite{zhang2005universal,zhang2005yang}. Note that the
$\check R_{I}(\pi/4)$ gate is also the braid gate defined in Eq. (\ref{eq:braid_gate}). Since we know that the gate $\check R_{I}(\theta)$ is not local equivalent to the CNOT gate if $\theta \neq \pi/4$, then the realization of the gate $\check R_{I}(\theta)$ via two CNOTs is optimal.

Consider the entangling power of the $\check R_{I}(\theta)$ gate. On the basis of Eq. (\ref{eq:e_p}), we know that 
\begin{equation}
    e_p\left(\check R_{I}(\theta)\right) = \frac 2 9 \sin^2(2\theta),
\end{equation}
which is maximal at $\theta = \pi/4$ or $\theta = 3\pi/4$ (local equivalent to CNOT gate). Note that the entangling power of the Yang-Baxter gates is completely determined by its eigenvalues \cite{padmanabhan2021local}. Apply Theorem \ref{theorem_generate_maximal} to gate $\check R_{I}(\theta)$. Then we know that it is a perfect entangle only when $\theta = \pi/4$ or $\theta = 3\pi/4$.

Based on the above decompositions of the $\check R_{I}(\theta)$ gate, it has three nature realizations on quantum computers. First realization is from two CNOTs based on Eq. (\ref{eq:R_I_two_CNOTs}). The second is from one $R_{zz}(2\theta)$ gate based on Eq. (\ref{eq:R_I_one_Rzz}). The third realization is given by direct pulse construction, which realizes the $R_{zz}(2\theta)$ gate from the rescaled cross-resonance operations. We apply the above three realizations on IBM quantum machine {\fontfamily{qcr}\selectfont ibmq\_montreal}. Specific details about the backend {\fontfamily{qcr}\selectfont ibmq\_montreal} can be found in Appendix \ref{App:B}. We apply process tomography to reconstruct the operation \cite{mohseni2008quantum}. For the process tomography, each qubit is prepared in the state $|0\rangle$, $|1\rangle$, $(|0\rangle+|1\rangle)/\sqrt 2$, and $(|0\rangle+i|1\rangle)/\sqrt 2$. Then measure in the Pauli $\sigma_x$, $\sigma_y$, and $\sigma_z$ bases. Therefore, the two-qubit gate tomography requires 144 circuits. Then we compute the average gate fidelity defined as \cite{horodecki1999general}
\begin{equation}
\label{eq:gate_fidelity}
    \mathcal F(\mathcal E,U) = \int d\psi \langle \psi|U^\dag \mathcal E(|\psi\rangle\langle\psi|)U|\psi\rangle,
\end{equation}
with the ideal unitary gate $U$. It can be easily computed using the Choi matrix of quantum operation \cite{nielsen2002simple}.

Since the CNOT gate is the only native two-qubit gate for IBM Quantum processors, the $R_{zz}$ gate is complied to the two-CNOT realization. Therefore, the CNOT and $R_{zz}$ gate realizations for the $\check R_{I}(\theta)$ gate are equivalent on IBM quantum processors. See FIG. \ref{fig_YBG_fidelity_1}. We test the tomography circuits on both {\fontfamily{qcr}\selectfont ibmq\_montreal} simulator and real {\fontfamily{qcr}\selectfont ibmq\_montreal} machine. The simulator supports pulse simulation. As expected, the pulse length of direct pulse realization is shorter compared to that of CNOT or $R_{zz}$ gate realization. See Appendix \ref{App:B} for detailed pulses. Accordingly, the pulse realization always has higher gate fidelity. The pulse length of direct pulse realization depends on the spectral parameter $\theta$, while it is a constant for the CNOT or the $R_{zz}$ gate realization. On the real device, the gate fidelity oscillates according to the pulse length. However, the simulator does not reflect on such a feature, since the simulator only adds homogeneous errors after each cross-resonance operation. We also calculate the error reduction of the pulse realization relative to the CNOT or the $R_{zz}$ realization, given by $(\mathcal F_\text{pulse}-\mathcal F_\text{CNOT})/(1-\mathcal F_\text{CNOT})$. Error reduction ranges from $13.1\%$ to $32.3\%$.

\subsection{\label{subsec:YBG_second}Realizations of second type of Yang-Baxter gate}

Up to an overall factor, the Yang-Baxter gate $\check R_{II}(\phi)$ defined in Eq. (\ref{eq:YBG_second}) has the exponential form
\begin{equation}
\label{eq:YBG_II_exp}
    \check R_{II}(\phi) \cong e^{i\frac \phi 2\left(\sigma_x\otimes \sigma_x \right)}e^{i\frac \phi 2\left(\sigma_y\otimes \sigma_y \right)}e^{i\frac \phi 2\left(\sigma_z\otimes \sigma_z \right)}.
\end{equation}
Applying the Clifford gate identities \cite{nielsenQuantumComputationQuantum2010}
\begin{equation}
    \sigma_x = H\sigma_zH; \quad \sigma_y = \sqrt{\sigma_x}^\dag\sigma_z\sqrt{\sigma_x},
\end{equation}
naturally gives the realization of $\check R_{II}(\phi)$ gate via three $R_{zz}$ gates, namely
\begin{widetext}
\begin{equation}
\label{eq:YBG_II_three_rzz}
    \Qcircuit @C=0.7em @R=1.5em {
    &&&&& \gate{H} & \multigate{1}{R_{zz}(\phi)} & \gate{H} & \gate{\sqrt{\sigma_x}} & \multigate{1}{R_{zz}(\phi)} & \gate{\sqrt{\sigma_x}^\dag} & \multigate{1}{R_{zz}(\phi)} & \qw & \qw \\
    \raisebox{1.2cm}{$\check R_{II}(\phi)=$} &&&&& \gate{H} & \ghost{R_{zz}(\phi)} & \gate{H} & \gate{\sqrt{\sigma_x}} & \ghost{R_{zz}(\phi)} & \gate{\sqrt{\sigma_x}^\dag} & \ghost{R_{zz}(\phi)} & \qw & \qw 
    }
\end{equation}
\end{widetext}
Here $\sqrt{\sigma_x}$ gate is equivalent to the gate $R_x(\pi/2)$, namely $\sqrt{\sigma_x} = e^{i\pi/4}R_x(\pi/2)$. Note that $\sqrt{\sigma_x}$ is a native single-qubit gate of IBM quantum processors \cite{IBMQ}. According to Theorem \ref{theorem_minimal_rzz}, we know that the above realization in terms of the $R_{zz}$ gate is optimal. 

\begin{figure}
    \begin{center}
    	\includegraphics[width=\columnwidth]{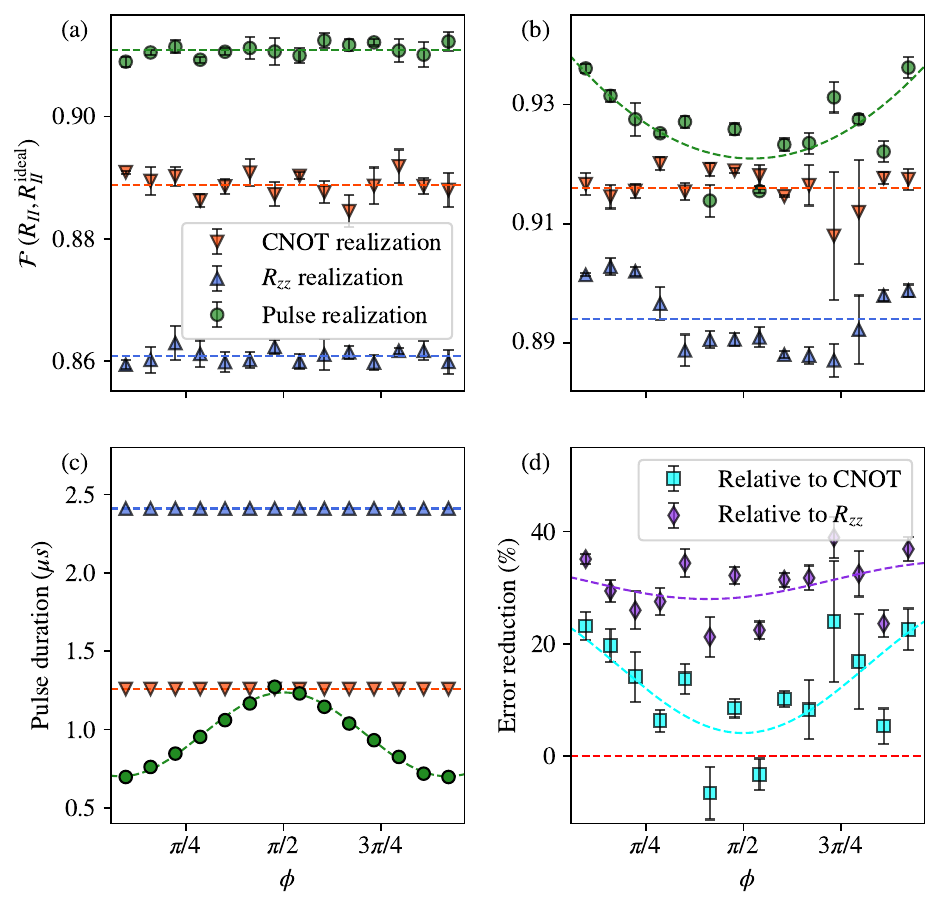}
    	\caption{Realizations of Yang-Baxter gate $\check R_{II}(\phi)$ on the quantum computer {\fontfamily{qcr}\selectfont ibmq\_montreal}. Measured gate fidelity on (a) FakeMontreal ({\fontfamily{qcr}\selectfont ibmq\_montreal} simultor) or (b) {\fontfamily{qcr}\selectfont ibmq\_montreal}. (c) The pulse duration of three different realizations. (d) The error reduction of pulse realization compared to the CNOT or $R_{zz}$ realization. The dashed line is the fitting. Each tomography circuit repeats four times and each time runs 4096 shots. }\label{fig_YBG_fidelity_2}
    \end{center}
\end{figure}

If we decompose the $R_{zz}$ gate into two CNOTs, that is, Eq. (\ref{eq:Rz_two_cnot}), we have a realization of the $\check R_{II}(\phi)$ gate with six CNOTs. According to \cite{zhang2004optimal,vidal2004universal,vatan2004optimal}, six CNOTs are not optimal. After applying Clifford gate identities, we can reformulate the RHS of Eq. (\ref{eq:YBG_II_three_rzz}) as
\begin{equation}
\label{eq:YBG_II_three_cnot}
    \Qcircuit @C=0.6em @R=1.5em {
    & \ctrl{1} & \gate{H} & \gate{R_z\left(\phi+\frac \pi 2\right)} & \ctrl{1} & \gate{H} & \ctrl{1} & \gate{\sqrt{\sigma_x}^\dag} & \qw \\
    & \targ & \gate{R_z(\phi)} & \qw & \targ & \gate{R^\dag_z(\phi)} & \targ & \gate{\sqrt{\sigma_x}} & \qw
    }
\end{equation}
The detailed derivation can be found in Appendix \ref{App:A}. Since the Yang-Baxter gate $\check R_{II}(\phi)$ is not an element of $SO(4)$, we know that three CNOTs are minimal.

The exponential form of $\check R_{II}(\phi)$ shown in Eq. (\ref{eq:YBG_II_exp}) gives the nonlocal parameters $\check R_{II}(\phi) = [\phi,\phi,\phi]$ if $\phi\leq \pi/2$ or $\check R_{II}(\phi) = [\phi,\pi-\phi,\pi-\phi]$ if $\phi> \pi/2$. In the geometric picture, it is the line $OA_3$ and $A_3A_1$. See FIG. \ref{fig_two-qubit}. When $\phi = \pi/2$, $\check R_{II}(\pi/2)$ is local equivalent to the permutation gate $[P] = [\pi/2,\pi/2,\pi/2]$. We can find that the Yang-Baxter gate $\check R_{II}(\phi)$ has the entangling power
\begin{equation}
    e_p\left(\check R_{II}(\phi)\right) = \frac 1 6\sin^2(2\phi).
\end{equation}
When $\phi = \pi/4$ or $\phi = 3\pi/4$, it has the maximal entangling power $1/6$, which is smaller than the maximal entangling power of the first type of Yang-Baxter gate $\check R_{I}(\theta)$. Based on Theorem \ref{theorem_generate_maximal}, we know that $\check R_{II}(\pi/4)$ and $\check R_{II}(3\pi/4)$ are perfect entanglers. For other spectral parameters $\phi$, the second type of Yang-Baxter gate cannot generate the maximal entangled states from the product states.

We test the three realizations of $\check R_{II}(\phi)$ on IBM quantum processor {\fontfamily{qcr}\selectfont ibmq\_montreal}, namely the three-$R_{zz}$ realization, three-CNOT realization, and the direct pulse realization. See FIG. \ref{fig_YBG_fidelity_2}. Qiskit compiles the $R_{zz}$ gate into two CNOTs. However, it can not compile to the three-$R_{zz}$ realization to the optimal three-CNOT realization. Therefore, the three-$R_{zz}$ realization has the longest circuit or pulses among the three realizations. Note that the direct pulse realization realizes the $R_{zz}$ gate from the rescaled cross-resonance operation. See Appendix \ref{App:B} for detailed pulses. Similarly to the simulation results for $\check R_{I}(\theta)$, the {\fontfamily{qcr}\selectfont ibmq\_montreal} simulator can not reflect the fidelity curve of the pulse realization. On the real processor, the pulse realizations always have higher gate fidelity except when $\phi$ is around $\pi/2$. When $\phi=\pi/2$, the pulse of $R_{zz}$ gate is equivalent to CNOT, therefore, providing no advantage. Compared to the three-$R_{zz}$ and three-CNOT realizations, the pulse realization has a maximal error reduction $38.9\%$ and $24.0\%$, respectively. 

\section{\label{sec:sim_YBE} Simulations of the Yang-Baxter equation}

In this section, we test the Yang-Baxter equation via the different realizations of the Yang-Baxter gates. Yang-Baxter equations given by the first and second type Yang-Baxter gates are presented in Sec. \ref{subsec:YBE_type_I} and \ref{subsec:YBE_type_II} respectively. 

\subsection{\label{subsec:YBE_type_I}Yang-Baxter equation from first type Yang-Baxter gate}

\begin{figure*}
    \begin{center}
    	\includegraphics[width=\textwidth]{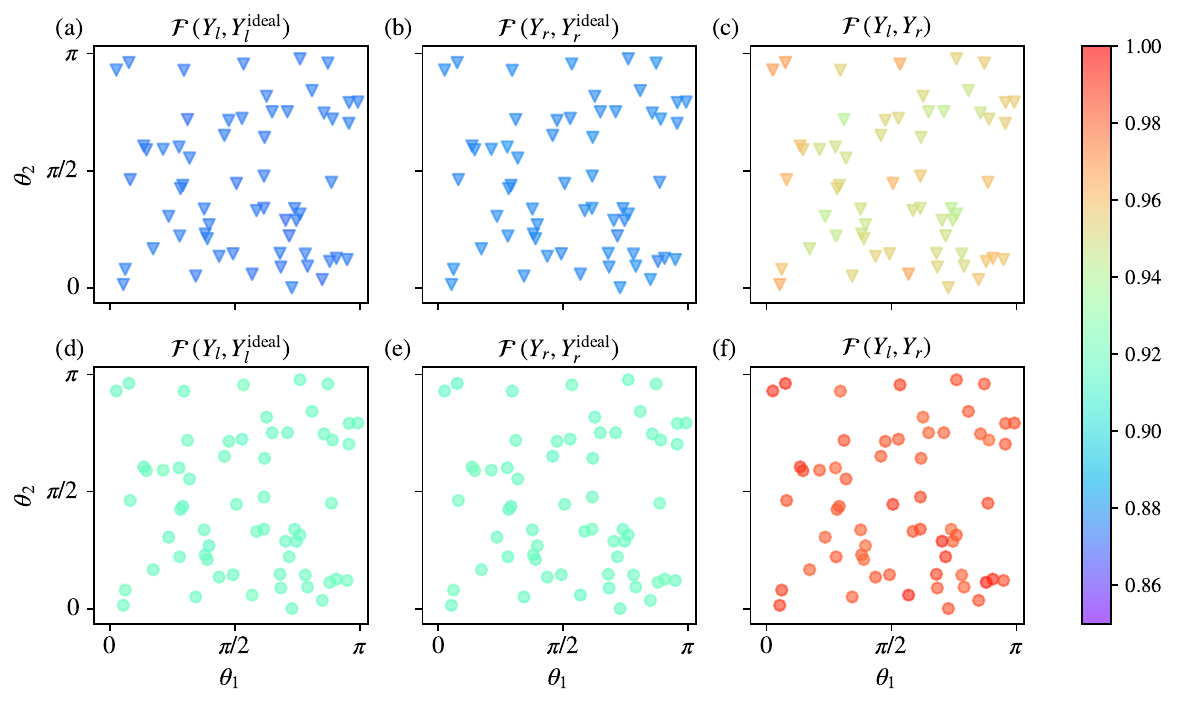}
    	\caption{Fidelities of Yang-Baxter equation from the first type of Yang-Baxter gate $\check R_I(\theta)$. Parameters $\theta_1$ and $\theta_2$ are randomly chosen. The three-qubit operations $Y_l$ and $Y_r$ are the LHS and RHS of the Yang-Baxter equation, respectively. The Yang-Baxter gate $\check R_I(\theta)$ is realized by (a)(b)(c) two CNOTs or (d)(e)(f) the rescaled pulse of the gate $R_{zz}$. Each tomography circuit runs 4096 shots. The results are obtained from qubits 24, 25, and 26 of the {\fontfamily{qcr}\selectfont ibmq\_montreal} simultor.}\label{fig_YBE_fidelity_1}
    \end{center}
\end{figure*}

Yang-Baxter equation has been experimentally verified on the basis of different proposals \cite{hu2008optical,zheng2013direct,vind2016experimental,zheng2018duality,wang2020experimental}. To verify the Yang-Baxter equation with the matrix $4\times 4$ $\check R$, an interference method has been applied in \cite{zheng2018duality,wang2020experimental}. However, the interference method only verifies the necessary condition for the Yang-Baxter equation. As pointed out in \cite{wang2020experimental}, quantum process tomography is required to fully verify the Yang-Baxter equation. 

Separately denote the LHS and RHS of Yang-Baxter equation as
\begin{subequations}
\begin{align}
    Y_l(\theta_1,\theta_2) = & \check R_{12}(\theta_1)\check R_{23}(\theta_{3})\check R_{12}(\theta_2); \\
    Y_r(\theta_1,\theta_2) = & \check R_{23}(\theta_2)\check R_{12}(\theta_{3})\check R_{23}(\theta_1).
\end{align}
\end{subequations}
Note that the angle $\theta_3$ is dependent on $\theta_1$ and $\theta_2$. Consider the first type Yang-Baxter gate defined in Eq. (\ref{eq:YBG_first}). Based on the parameter $\tan(\theta-\pi/4) = \mu$, we know that
\begin{equation}
\label{eq:theta_I}
    \tan\theta_3 = \frac{\sin(\theta_1+\theta_2)}{\cos(\theta_1-\theta_2)}.
\end{equation}
We separately perform process tomography on $Y_l(\theta_1,\theta_2)$ and $Y_r(\theta_1,\theta_2)$. Then calculate its average gate fidelity defined in Eq. (\ref{eq:gate_fidelity}). The three-qubit process tomography requires total $4^3\times 3^3 = 1728$ circuits. Therefore, given a pair of parameter $(\theta_1,\theta_2)$, more than 3400 circuits are required to verify the Yang-Baxter equation. Instead of real quantum processors, we test the Yang-Baxter equation on backend simulator, which has the error parameters closed to the real machine. 

We randomly choose the parameters $\theta_1,\theta_2$ and determine $\theta_3$ according to the relation (\ref{eq:theta_I}). See FIG. \ref{fig_YBE_fidelity_1} for the Yang-Baxter equation fidelity on {\fontfamily{qcr}\selectfont ibmq\_montreal} simultor. We also calculate the gate fidelity of $Y_l(\theta_1,\theta_2)$ and $Y_r(\theta_1,\theta_2)$ relative to the ideal gate. In the two-CNOT realization of $\check R_I$, the average of $\mathcal F(Y_l,Y_l^\text{ideal})$ and $\mathcal F(Y_r,Y_r^\text{ideal})$ are $87.4\%$ and $87.7\%$ respectively. In the pulse realization, the average of $\mathcal F(Y_l,Y_l^\text{ideal})$ and $\mathcal F(Y_r,Y_r^\text{ideal})$ are $91.9\%$ and $91.8\%$ respectively. The fidelity of $Y_l$ and $Y_r$ decay in a similar degree. Then it is expected that $\mathcal F(Y_l,Y_r)$ is higher than $\mathcal F(Y_l,Y_l^\text{ideal})$ and $\mathcal F(Y_r,Y_r^\text{ideal})$. The pulse realization gives the Yang-Baxter equation fidelity $\mathcal F(Y_l,Y_r)$ close to 1 (around $98.5\%$) even though the gate operation is noisy. We do not show the fidelity from the one-$R_{zz}$ realization, since it is equivalent to the two-CNOT realization on IBM quantum processors.

\subsection{\label{subsec:YBE_type_II}Yang-Baxter equation from second type Yang-Baxter gate}

\begin{figure*}
    \begin{center}
    	\includegraphics[width=\textwidth]{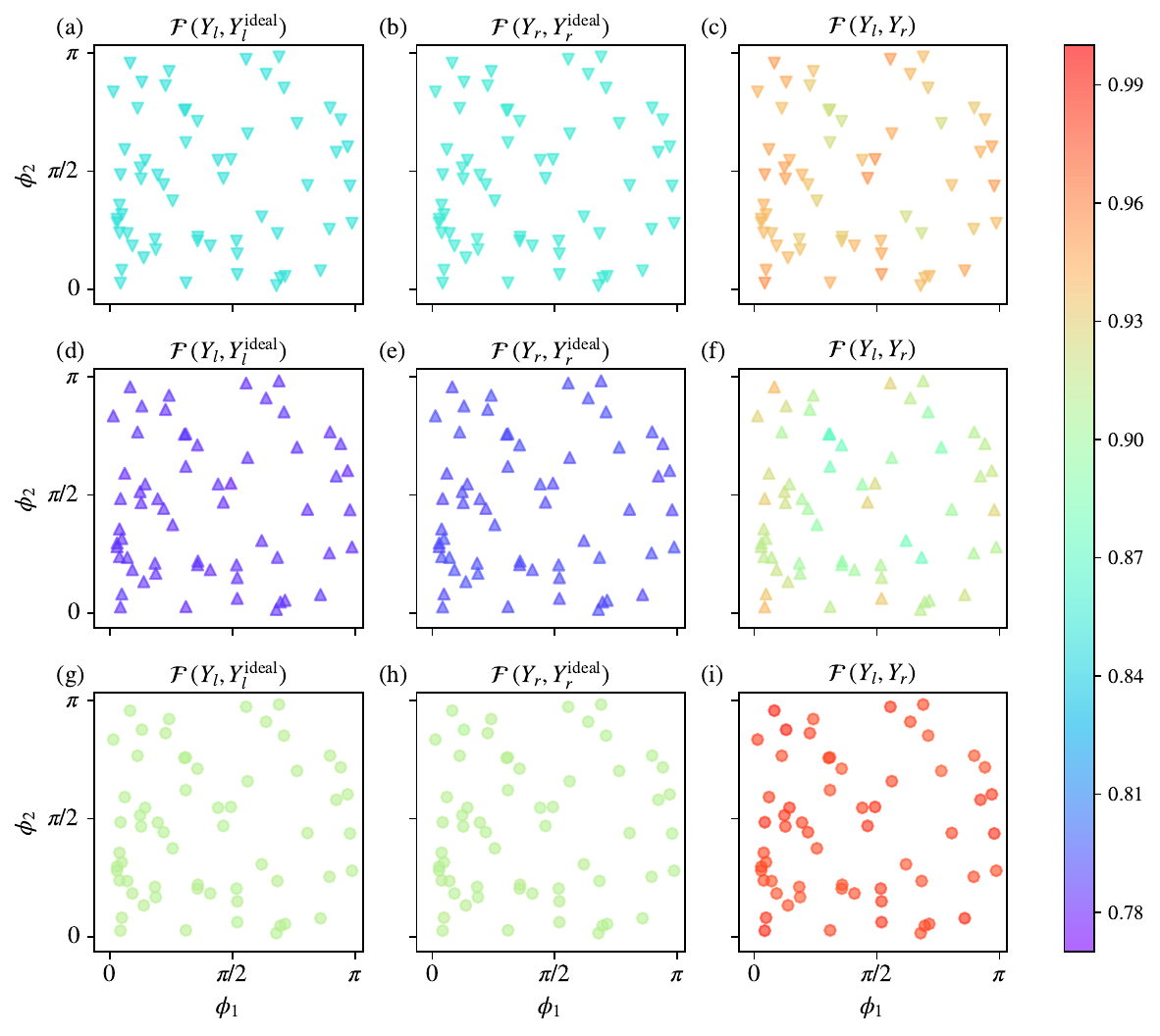}
    	\caption{Fidelities of Yang-Baxter equation from the first type of Yang-Baxter gate $\check R_{II}(\phi)$. Parameters $\theta_1$ and $\theta_2$ are randomly chosen. The Yang-Baxter gate $\check R_I(\theta)$ is realized by (a)(b)(c) three CNOTs, (d)(e)(f) three $R_{zz}$ gates, or (g)(h)(i) the direct pulse constructions. Each tomography circuit runs 4096 shots. The results are obtained from qubits 24, 25 and 26 of the {\fontfamily{qcr}\selectfont ibmq\_montreal} simultor. }\label{fig_YBE_fidelity_2}
    \end{center}
\end{figure*}

Consider the Yang-Baxter equation from the second type of Yang-Baxter gate $\check R_{II}(\phi)$ defined in Eq. (\ref{eq:YBG_second}). The parameter $\tan\phi=\mu$ gives the relation
\begin{equation}
    \tan\phi_3 = \tan\phi_1+\tan\phi_2.
\end{equation}
We randomly choose the parameters $\phi_1$ and $\phi_2$. Then run the process tomography on different realizations of $Y_l(\phi_1,\phi_2)$ and $Y_r(\phi_1,\phi_2)$. See FIG. \ref{fig_YBE_fidelity_2}. In the three-$R_{zz}$ realization, the average of $\mathcal F(Y_l,Y_l^\text{ideal})$ and $\mathcal F(Y_r,Y_r^\text{ideal})$ are $78.4\%$ and $\%79.2$ respectively. The Yang-Baxter equation from the pulse realization maintains a relatively high fidelity. The average values of $\mathcal F(Y_l,Y_l^\text{ideal})$ and $\mathcal F(Y_r,Y_r^\text{ideal})$ are around $91.3\%$. It is interesting to see that even though the fidelity of $Y_r$ and $Y_l$ decays in a similar way, the fidelity of the Yang-Baxter equation $\mathcal F(Y_l,Y_r)$ is subject to fluctuations. For example, see FIG. \ref{fig_YBE_fidelity_2} (c) and (f). The pulse realization gives the Yang-Baxter equation fidelity $\mathcal F(Y_l,Y_r)$ close to 1 (the average value is $98.1\%$). It implies that if we replace a pair of Yang-Baxter gate of the integrable circuit based on the Yang-Baxter equation, the fidelity of whole circuit would not be significantly changed. Note that the errors are more complicated in real quantum processors. The Yang-Baxter equation may provide a method for benchmarking the correlated errors of quantum computers. We leave it for future study.

\section{\label{sec:conclusion}Conclusions}

In this study, we have shown how to realize two types of Yang-Baxter gates through gate decompositions. We argue that the first type Yang-Baxter gate $\check R_I$ requires at most two CNOTs or one $R_{zz}$ gate. When the spectral parameter $\theta = \pi/4$, it is local equivalent to the CNOT gate. In terms of the second type Yang-Baxter gate $\check R_{II}$, it can be realized from three CNOTs or three $R_{zz}$ gates. Based on the geometric picture of the two-qubit gate \cite{zhang2003geometric,zhang2004optimal}, we argue that these realizations are optimal (in terms of the number of CNOT or $R_{zz}$ gate). Moreover, we have shown how to systematically realize these two types of Yang-Baxter gates via pulse control, specified for IBM quantum processors \cite{alexander2020qiskit}. The pulse realization of the Yang-Baxter gate has the highest gate fidelity among other realizations. We also show the verification of the Yang-Baxter equation on the simulator of IBM quantum processor. The results show that the Yang-Baxter equation fidelity approaches 1 with noisy operation. It suggests the errors to be acting homogeneously on the qubits. Both sides of the Yang-Baxter equation deteriorate in a similar way. It would be interesting to explore benchmarking the correlated errors of quantum computers via the Yang-Baxter equation in the future. The Yang-Baxter gate is the building block of the integrable circuit \cite{vanicat2018integrable,miao2022floquet} and is closely related to the simulation of integrable systems \cite{keenan2022evidence,yu2023simulating}. Our work provides an experimental guide and standard for future studies on quantum simulation of integrable systems.

\begin{acknowledgments}

The authors thank helpful discussions with Ryo Suzuki. The work of K.Z. was supported by the National Natural Science Foundation of China under Grant Nos. 12305028 and 12247103, and the Youth Innovation Team of Shaanxi Universities. The work of K.H. was supported by the National Natural Science Foundation of China (Grant Nos. 12275214, 12247103, and 12047502), the Natural Science Basic Research Program of Shaanxi Province Grant Nos. 2021JCW-19 and 2019JQ-107, and Shaanxi Key Laboratory for Theoretical Physics Frontiers in China. This research used quantum computing resources of the Oak Ridge Leadership Computing Facility, which is a DOE Office of Science User Facility supported under Contract DE-AC05-00OR22725. This research is funded by the U.S. Department of Energy, Office of Science, National Quantum Information Science Research Centers, Co-Design Center for Quantum Advantage under Contract No. DE-SC0012704. 

\end{acknowledgments}

\appendix

\section{\label{App:A} Three CNOT gates constructions of the Yang-Baxter gate}

We need the following circuit identities to derive the gate decomposition (\ref{eq:YBG_II_three_cnot}).
\begin{subequations}
\begin{equation}
  \label{eq:circuit_identity_1}\Qcircuit @C=0.7em @R=1.5em {
    & \gate{H} & \gate{\sqrt{\sigma_x}} & \gate{H} & \qw 
    } \hspace{1mm} = \hspace{3.3mm}
    \Qcircuit @C=0.7em @R=1.5em {
    & \gate{S} & \qw 
    }\vspace{0.1cm}  
\end{equation}
\begin{equation}
    \label{eq:circuit_identity_2}\Qcircuit @C=0.7em @R=1.5em {
    & \gate{H} & \ctrl{1} & \gate{H} & \qw \\
    & \gate{H} & \targ & \gate{H} &\qw & 
    } \hspace{1mm} \raisebox{-0.6cm}{$=$} \hspace{3.3mm}
    \Qcircuit @C=0.7em @R=1.5em {
    & \targ & \qw \\
    & \ctrl{-1} & \qw 
    }\vspace{0.1cm}
\end{equation}
\begin{equation}
    \label{eq:circuit_identity_3}\Qcircuit @C=0.7em @R=1.5em {
    & \qw & \ctrl{1} & \qw \\
    & \gate{R_x(\theta)} & \targ & \qw & 
    } \hspace{1mm} \raisebox{-0.6cm}{$=$} \hspace{3.3mm}
    \Qcircuit @C=0.7em @R=1.5em {
    & \ctrl{1} & \qw & \qw \\
    & \targ & \gate{R_x(\theta)} & \qw 
    }\vspace{-0.1cm}
\end{equation}
\begin{equation}
    \label{eq:circuit_identity_4}\Qcircuit @C=0.7em @R=1.5em {
    & \gate{R_z(\theta)} & \ctrl{1} & \qw \\
    & \qw & \targ & \qw & 
    } \hspace{1mm} \raisebox{-0.6cm}{$=$} \hspace{3.3mm}
    \Qcircuit @C=0.7em @R=1.5em {
    & \ctrl{1} & \gate{R_z(\theta)} & \qw \\
    & \targ & \qw & \qw 
    }\vspace{-0.1cm}
\end{equation}
\begin{equation}
    \label{eq:circuit_identity_5}\Qcircuit @C=0.7em @R=2em {
    & \ctrl{1} & \qw & \ctrl{1} & \qw \\
    & \targ & \gate{S} & \targ & \qw 
    }\hspace{3mm} \raisebox{-0.6cm}{$=$} \hspace{3.3mm}
    \Qcircuit @C=0.7em @R=1.5em {
    & \qw & \ctrl{1} & \gate{S} & \qw & \qw \\
    & \gate{H} & \targ & \gate{H} & \gate{S} & \qw  
    }
\end{equation}
\end{subequations}
Note that the identities (\ref{eq:circuit_identity_3}) and (\ref{eq:circuit_identity_4}) hold for arbitrary angle $\theta$. The derivation of gate decomposition (\ref{eq:YBG_II_three_cnot}) is shown below.
\begin{widetext}
    \begin{align}
    \check R_{II}(\phi) =& \hspace{4mm} \Qcircuit @C=0.7em @R=1.5em {
    & \ctrl{1} & \qw & \ctrl{1} & \gate{H} & \ctrl{1} & \qw & \ctrl{1} & \gate{H} & \gate{\sqrt{\sigma_x}} & \ctrl{1} & \qw & \ctrl{1} & \gate{\sqrt{\sigma_x}^\dag} & \qw \\
    & \targ & \gate{R_z(\theta)} & \targ & \gate{H} & \targ & \gate{R_z(\theta)} & \targ & \gate{H} & \gate{\sqrt{\sigma_x}} & \targ & \gate{R_z(\theta)} & \targ & \gate{\sqrt{\sigma_x}^\dag} & \qw  
    }\nonumber \\
    =& \hspace{4mm}\Qcircuit @C=0.7em @R=1.5em {
    & \ctrl{1} & \qw & \gate{H} & \targ & \ctrl{1} & \gate{S} & \ctrl{1} & \targ & \gate{H} & \qw & \qw & \ctrl{1} & \gate{\sqrt{\sigma_x}^\dag} & \qw \\
    & \targ & \gate{R_z(\theta)} & \gate{H} & \ctrl{-1} & \targ & \gate{R_z(\theta)} & \targ & \ctrl{-1} & \gate{H} & \gate{\sqrt{\sigma_x}} & \gate{R_z(\theta)} & \targ & \gate{\sqrt{\sigma_x}^\dag} & \qw   
    }\nonumber \\
    =& \hspace{4mm}\Qcircuit @C=0.7em @R=1.5em {
    & \ctrl{1} & \qw & \gate{H} & \ctrl{1} & \gate{R_z(\theta)} & \ctrl{1} & \gate{H} & \qw & \qw & \ctrl{1} & \gate{\sqrt{\sigma_x}^\dag} & \qw \\
    & \targ & \gate{R_z(\theta)} & \gate{H} & \targ & \gate{S} & \targ & \gate{H} & \gate{\sqrt{\sigma_x}} & \gate{R_z(\theta)} & \targ & \gate{\sqrt{\sigma_x}^\dag} & \qw 
    }\nonumber \\
    =& \hspace{4mm}\Qcircuit @C=0.7em @R=1.5em {
    & \ctrl{1} & \gate{H} & \gate{R_z(\theta)} & \qw & \ctrl{1} & \gate{S} & \gate{H} & \qw & \qw & \qw & \ctrl{1} & \gate{\sqrt{\sigma_x}^\dag} & \qw \\
    & \targ & \gate{R_z(\theta)} & \gate{H} & \gate{H} & \targ & \gate{H} & \gate{S} & \gate{H} & \gate{\sqrt{\sigma_x}} & \gate{R_z(\theta)} & \targ & \gate{\sqrt{\sigma_x}^\dag} & \qw 
    }\nonumber \\
    =& \hspace{4mm}\Qcircuit @C=0.7em @R=1.5em {
    & \ctrl{1} & \gate{H} & \gate{R_z\left(\theta+\frac \pi 2\right)} & \ctrl{1} & \gate{H} & \ctrl{1} & \gate{\sqrt{\sigma_x}^\dag} & \qw \\
    & \targ & \gate{R_z(\theta)} & \qw & \targ & \gate{R^\dag_z(\theta)} & \targ & \gate{\sqrt{\sigma_x}} & \qw 
    }
\end{align}
\end{widetext}
The first line comes from decomposing the $R_{zz}(\theta)$ gate into two CNOTs. The second line is obtained by moving the Hadamard gates across the CNOT. The relations (\ref{eq:circuit_identity_1}), (\ref{eq:circuit_identity_2}), (\ref{eq:circuit_identity_3}), and (\ref{eq:circuit_identity_4}) are applied. The third line comes from rewriting the two CNOTs into one SWAP gate and one CNOT. Then the two SWAPs cancel. The fourth line is obtained by applying the relation (\ref{eq:circuit_identity_4}). The last line comes from merging the single-qubit gates. Note that we have $R^\dag_z(\theta) = \sigma_xR_z(\theta)\sigma_x$ and $\sqrt{\sigma_x} = \sigma_x\sqrt{\sigma_x}^\dag$.

\section{\label{App:B} Details on the IBM quantum processor and the pulse realizations}

\begin{figure}
    \begin{center}
    	\includegraphics[width=\columnwidth]{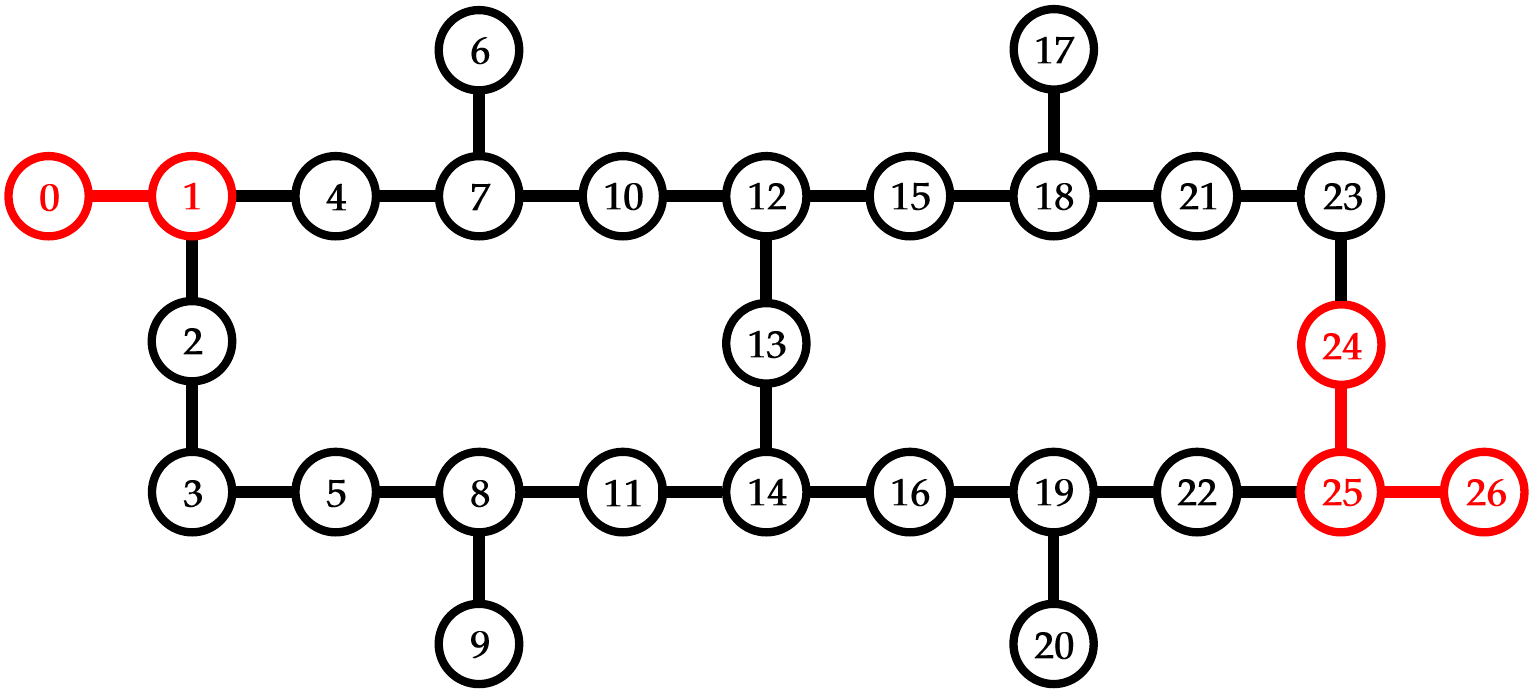}
    	\caption{The qubit layout of {\fontfamily{qcr}\selectfont ibmq\_montreal}. We realize the Yang-Baxter gate on qubit 0 and 1. When simulating Yang-Baxter equation, we use the qubit 24, 25, and 26 of the {\fontfamily{qcr}\selectfont ibmq\_montreal} simulator.}\label{fig_ibmq_montreal}
    \end{center}
\end{figure}

\begin{figure}
    \begin{center}
    	\includegraphics[width=\columnwidth]{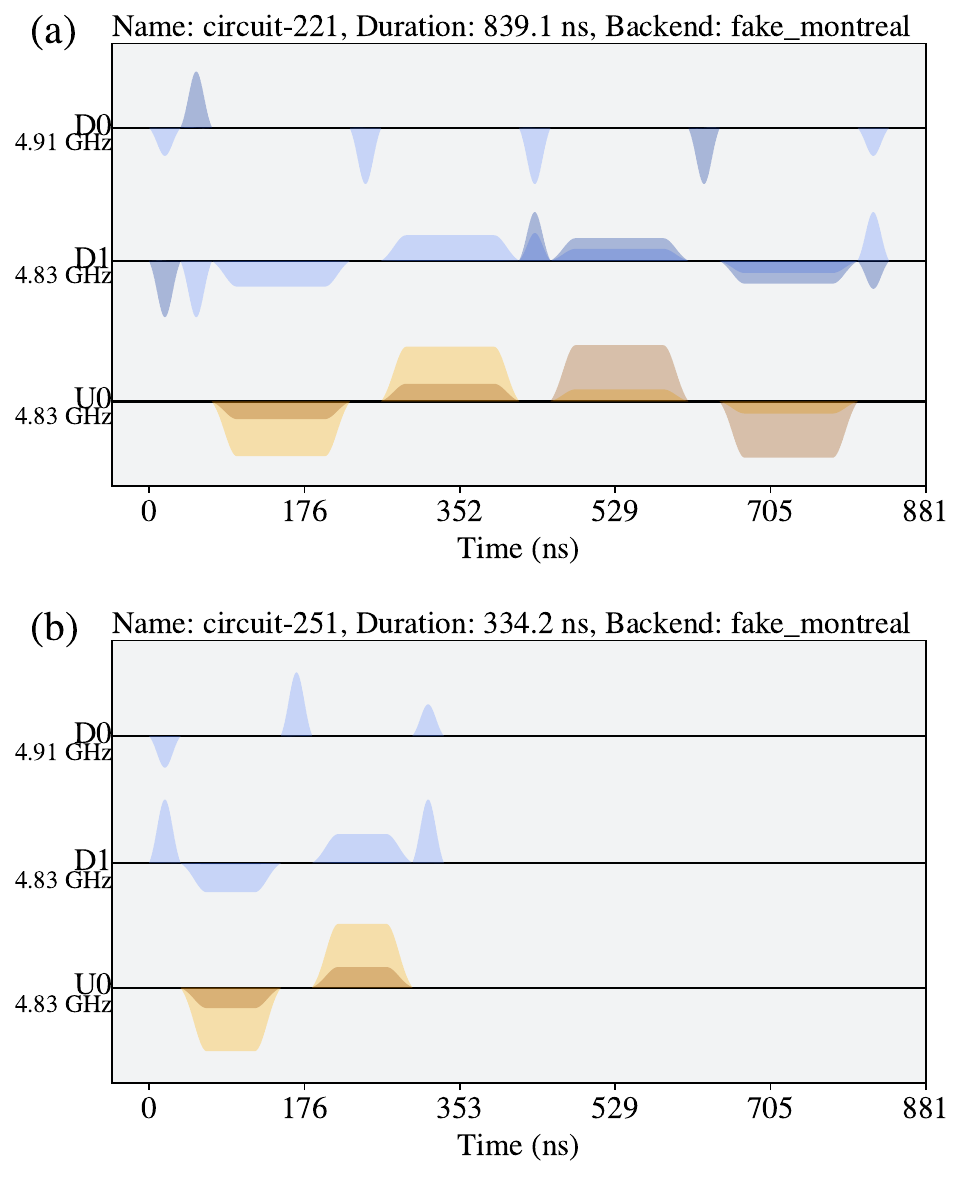}
    	\caption{The pulses of Yang-Baxter gate $\check R_I(\pi/3)$. (a) The two-CNOT realization given by Eq. (\ref{eq:R_I_two_CNOTs}). (b) Realized from one pair of rescaled cross-resonance operations. The virtual $z$-axis rotation is omitted here for simplicity.  }\label{fig_YBG_I_pulse}
    \end{center}
\end{figure}

\begin{figure}
    \begin{center}
    	\includegraphics[width=\columnwidth]{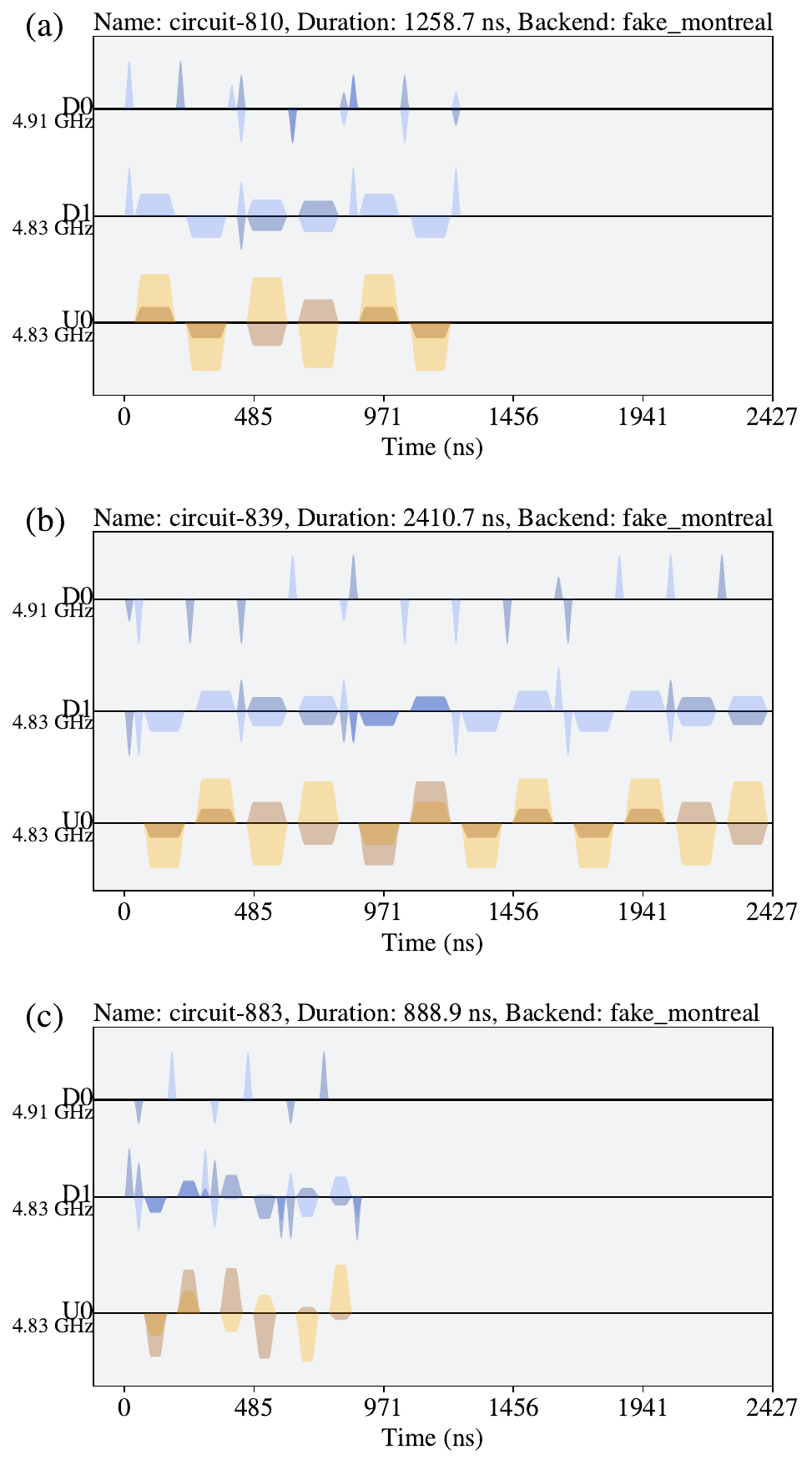}
    	\caption{The pulses of Yang-Baxter gate $\check R_{II}(\pi/4)$. (a) The three-CNOT gate realization given by Eq. (\ref{eq:YBG_II_three_cnot}). (b) The three-$R_{zz}$ gates realization given by Eq. (\ref{eq:YBG_II_three_rzz}). (c) Realized from three pairs of rescaled cross-resonance operations. The virtual $z$-axis rotation is omitted here for simplicity.  }\label{fig_YBG_II_pulse}
    \end{center}
\end{figure}

The {\fontfamily{qcr}\selectfont ibmq\_montreal} processor has 27 qubits. It has Quantum Volume 128. See FIG. \ref{fig_ibmq_montreal} for the qubit layout. Process tomography is a heavy resource-consuming task. The data of FIG. \ref{fig_YBG_fidelity_1} was collected from April 4th to April 8th. The data of FIG. \ref{fig_YBG_fidelity_2} was collected from March 17th to March 20th. We choose qubits 0 and 1 in order to mitigate the cross-talk errors between qubits. IBM quantum processors are calibrated regularly. Therefore, the benchmark metrics varied during data collection. Typically, read-out errors are around $2\%$. The infidelity of the CNOT gate for qubits 0 and 1 is less than $1\%$. The infidelity of the single-qubit gate $\sqrt{\sigma_x}$ is in the order of $0.01\%$. The relaxation time $T_1$ and $T_2$ are above $100\mu s$.

We present the pulses of different realizations of Yang-Baxter gates $\check R_{I}(\theta)$ and $\check R_{II}(\phi)$ in FIGs. \ref{fig_YBG_I_pulse} and \ref{fig_YBG_II_pulse}. For the first type Yang-Baxter gate $\check R_{I}(\theta)$, the pulse length of the direct pulse realization (rescaled cross-resonance operation) is less than half of the pulse length of the two-CNOT realization. For the second type Yang-Baxter gate $\check R_{II}(\phi)$, the pulse length is around one third compared to the three-$R_{zz}$ realization.


\providecommand{\noopsort}[1]{}\providecommand{\singleletter}[1]{#1}%

\end{document}